\documentclass[
  aps,
  pra,
  twocolumn,
  superscriptaddress,
  nolongbibliography,
  nofootinbib,
  10pt,
]{revtex4-2}

\usepackage{graphics}
\usepackage{graphicx}
\usepackage{bbm,bm}
\usepackage{amsfonts}
\usepackage{amsmath}
\usepackage[dvipsnames]{xcolor}
\usepackage{nicefrac} 
\usepackage{siunitx}
\usepackage{braket}
\usepackage{bm}

\ifx\pdftexversion\undefined
\usepackage[
  dvips
]{hyperref}
\else
\usepackage[
]{hyperref}
\fi
\hypersetup{
	hidelinks,
	colorlinks=true,
	linkcolor=black,
	citecolor=MidnightBlue,
	urlcolor=MidnightBlue,
}
\newcommand{\kd}{\ensuremath{k_{\rm d}}}

\newcommand{\uc}{\ensuremath{u_{\rm c}}}
\newcommand{\uz}{\ensuremath{u_{\rm c, 0}}}
\newcommand{\he}{$^{4}\mathrm{He}^*$}

\newcommand{\dd}{\ensuremath{\mathrm{d}}}
\newcommand{\ee}{\ensuremath{\mathrm{e}}}
\newcommand{\ii}{\ensuremath{\mathrm{i}}}
\newcommand{\figref}[1]{Fig.~\ref{#1}}
\newcommand{\edfigref}[1]{Extended Data Fig.~\ref{#1}}

\newcounter{contbibcnt}

\makeatletter
\let\REVTEX@thebibliography\thebibliography
\let\REVTEX@endthebibliography\endthebibliography
\renewcommand{\thebibliography}[1]{%
  \@fileswfalse
  \REVTEX@thebibliography{#1}%
  \@fileswtrue
}
\renewcommand{\endthebibliography}{%
  \@fileswfalse
  \REVTEX@endthebibliography
  \@fileswtrue
}
\makeatother

\makeatletter
\def\maketitle{
\@author@finish
\title@column\titleblock@produce
\suppressfloats[t]}
\makeatother

\newif\ifsupplement
\supplementfalse

\begin{document}
\preprint{}

\title{
Observation of universal non-Gaussian statistics of the order parameter across a continuous phase transition
}

\newcommand{\LCF}{Universit\'e Paris-Saclay, Institut d'Optique Graduate School, CNRS, Laboratoire Charles Fabry, 91127, Palaiseau, France}
\newcommand{\PHLAM}{Univ. Lille, CNRS, UMR 8523 -- PhLAM -- Laboratoire de Physique
des Lasers Atomes et Mol\'ecules, F-59000 Lille, France}
\newcommand{\LPTMC}{Sorbonne Université, CNRS, Laboratoire de Physique Théorique de la Matière Condensée, LPTMC, F-75005 Paris, France}
\newcommand{\ENS}{Univ Lyon, Ens de Lyon, CNRS, Laboratoire de Physique, F-69342 Lyon, France}
\newcommand{\equalContribution}{These authors contributed equally to this work.}

\author{Maxime Allemand}
\thanks{\equalContribution}
\affiliation{\LCF}

\author{Géraud Dupuy}
\thanks{\equalContribution}
\affiliation{\LCF}

\author{Paul Paquiez}
\thanks{\equalContribution}
\affiliation{\LCF}

\author{Nicolas Dupuis}
\affiliation{\LPTMC}

\author{Adam Rançon}
\affiliation{\PHLAM}
\affiliation{Institut Universitaire de France}

\author{Tommaso Roscilde}
\affiliation{\ENS}

\author{Thomas Chalopin}
\affiliation{\LCF}

\author{David Clément}
\email[Electronic address: ]{david.clement@institutoptique.fr}%
\affiliation{\LCF}

\begin{abstract}
Second-order phase transitions are characterised by critical scaling and universality \cite{zinn-justin:2002}.
The singular behaviour of thermodynamic quantities at the transition, in particular, is determined by critical exponents of the universality class of the transition.
However, critical properties are also characterised by the probability distribution of the order parameter across the transition 
\cite{binder:1981a, bruce:1992}, where non-Gaussian statistics are expected \cite{bouchaud:1990, bramwell:1998, botet:2002}, but remain largely unexplored \cite{joubaud:2008}.
Here, making use of single-atom-resolved detection in momentum space \cite{cayla:2018a}, we measure the full probability distribution of the order-parameter amplitude across a continuous phase transition in an interacting lattice Bose gas \cite{greiner:2002}.
We find that fluctuations are captured by an effective potential ---  reconstructed from the measured probability distribution by analogy with Landau theory \cite{landau:1937} --- displaying a non-trivial minimum in the superfluid (ordered) phase, which vanishes at the transition point.
Additionally, we observe non-Gaussian statistics of the order parameter near the transition, distinguished by non-zero high-order cumulants undergoing abrupt sign changes.
We show numerically that these sign changes of the cumulants obey critical scaling in homogeneous systems, and that their experimental behaviour is not reproduced by classical models, whereas it is captured by a low-temperature quantum model.
Our results underscore the crucial role of order parameter statistics in probing critical phenomena and universality.
\end{abstract}

\maketitle

Continuous phase transitions driven by spontaneous symmetry breaking~\cite{landau:1937, ginzburg:1950} are fundamental phenomena covering all branches of physics \cite{nishimori:2010}.
The concept of a fluctuating order parameter $\psi$ plays an essential role in their description. Its  average amplitude $\braket{|\psi|}$ was shown by Landau~\cite{landau:1937} to distinguish the ordered phase, where it assumes a non-zero value in the thermodynamic limit, from the disordered phase where it vanishes. 
Landau's mean-field theory, however, neglects 
fluctuations in the order parameter and fails to capture the singular behaviour at the transition, {\it i.e.} the critical regime. 

\begin{figure}[!t]
\centering
\includegraphics[scale=1]{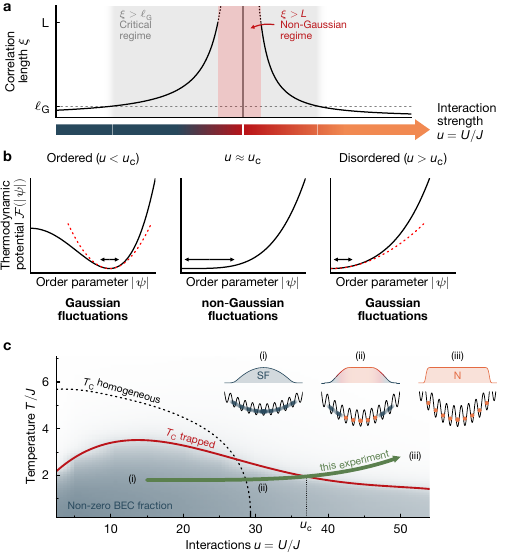}
\caption{
\textbf{Continuous phase transitions}.
\textbf{a.}
The critical regime (grey area) is reached when the correlation length $\xi$
exceeds the Ginzburg length $\ell_{\rm G}$. 
Closer to the transition point, $\xi$ exceeds the system size $L$ and non-Gaussian statistics are expected (red area).
\textbf{b.} The critical properties of the transition are described by an effective thermodynamic potential $\mathcal{F}(|\psi|)$ --- analogous to the Landau free energy --- that describes enhanced and non-Gaussian fluctuations of the order-parameter amplitude $|\psi|$ near the transition. 
\textbf{c.} We explore the superfluid-to-normal fluid phase transition of the Bose-Hubbard model, implemented by ultracold lattice bosons. 
A harmonic trap affects the interaction-dependent critical temperature $T_\mathrm{c}(u)$ between a superfluid (blue region, i) and a normal gas (finite-temperature Mott insulator, white region, iii).
Near the critical regime (ii), both phases coexist, with a normal core surrounded by superfluid shells.
}
\label{fig:fig1}
\end{figure}

These phenomena are in fact driven by the large fluctuations that develop as the system approaches the phase transition. As illustrated in \figref{fig:fig1}a, this aspect is captured by the divergence of the correlation length $\xi$ of the local order parameter, $\braket{\psi^{\dagger}(0) \psi(\bm{x})}_{\mathrm{c}} \sim \exp(-|\bm{x}|/\xi)$  \cite{bramwell:2001, nishimori:2010}. 
As this length grows beyond microscopic scales (e.g. the interparticle distance), it reaches the Ginzburg length $\ell_\mathrm{G}$, marking the onset of the critical regime. The latter is characterised by power-law singularities of thermodynamic quantities, exhibiting universal critical exponents.
These features are well understood in the framework of Wilson's renormalisation group \cite{wilson:1974, fisher:1998}. Systems that display the same large-scale behaviour near their critical points are classified in the same universality class (\emph{i.e.}, same critical exponents), despite having different microscopic physics.
In finite-size systems, though, a peculiar behaviour appears within the critical regime when the correlation length $\xi$ becomes larger than the linear system size $L$. While for $\xi <L$ the central limit theorem (CLT) implies that the order-parameter statistics is Gaussian, by contrast, when $\xi > L$, the CLT breaks down, and the order-parameter statistics become non-Gaussian \cite{bouchaud:1990, bramwell:1998, botet:2002}.

The probability distribution function (PDF) of the order parameter, which fully captures its statistical properties, is a fundamental quantity in the description of phase transitions.
At the transition point, its non-Gaussian shape is universal, \emph{i.e.} it depends on the universality class of the transition as well as global geometric properties (boundary conditions, aspect ratio between different dimensions, etc.). The PDF has received extensive theoretical and numerical attention over the past few decades 
\cite{binder:1981a, chen:1996, tsypin:2000, balog:2022, sahu:2025, rancon:2025},
 focusing in the thermodynamic limit where $\xi$ and $L$ are sent to infinity at fixed $\xi/L$. 
From an experimental perspective, however, measuring the non-Gaussian PDF  at criticality poses major challenges.
The non-Gaussian regime is  a narrow region of the control parameter near the transition point (\figref{fig:fig1}a) that shrinks with increasing system size.
In a macroscopic system with Avogadro number of particles, it reduces to the critical point, requiring therefore an unrealistic precision in the tuning of the control parameter.
Furthermore, characterising critical non-Gaussian statistics is inherently difficult
\cite{hsu:1995}, and, to our knowledge, has only been reported in a classical, mesoscopic system of liquid crystals~\cite{joubaud:2008, takeuchi:2010}.
More generally, non-Gaussian statistics are a diagnostic of complexity and criticality in physics and across many real-world phenomena \cite{sornette:2006}, and their characterisation is fundamental to the understanding of complex systems.

Modern experimental platforms implement quantum many-body models in which phase transitions with substantial non-Gaussian critical regimes are expected, thanks to small system sizes of tens to hundreds of particles~\cite{islam:2011, jurcevic:2017, makhalov:2019, xu:2020a, ebadi:2021, fontaine:2022, chen:2023, ho:2025}.
In these platforms, single-particle-resolved detection methods~\cite{ott:2016a} provide access to the full counting statistics of observables~\cite{bohnet:2016, schweigler:2017, ebadi:2021, chen:2023, herce:2023, joshi:2025}.
These advances enable the exploration of the universal order-parameter statistics, which is complementary to the standard measurements of critical exponents \cite{pelissetto:2002, donner:2007, zhang:2012, shao:2024}.

Here, we leverage these techniques to measure the distribution of the condensate order parameter across the superfluid transition, using an ultracold gas of $\sim 4000$ bosons in optical lattices (\figref{fig:fig1}c).
Our main results are threefold. 
First, we observe non-Gaussian order-parameter statistics in the critical regime near the transition, thereby providing a confirmation of renormalisation group predictions \cite{balog:2022}.
Second, we show that the measured PDFs are well approximated by a phenomenological approach similar in spirit to Landau's \cite{landau:1937}.
We additionally explain why this simple picture effectively describes our measurements.
Third, we characterise the non-Gaussian nature of the statistics through high-order cumulants of the order-parameter amplitude $| \psi |$ and we reveal abrupt sign changes of these cumulants across the transition.
We show numerically that these sign changes of cumulants exhibit critical scaling across the transition in homogeneous systems, and that their experimentally observed behaviour is not captured by classical models, while it is reproduced by a low-temperature quantum model with the same universal behaviour as that of lattice bosons.
The characterisation of the cumulants' behaviour stems from the unique level of diagnostics offered by quantum simulators. It represents a new perspective on the fundamental nature of critical fluctuations \cite{bramwell:1998,bertin:2005,coleman:2005}.
\newline

Our experiment investigates the superfluid-to-normal fluid transition in a lattice Bose gas.
We adiabatically load ultracold metastable helium (\he) atoms in a cubic optical lattice \cite{carcy:2021}, realising the Bose-Hubbard model (BHM) 
\begin{equation}
  \hat H = -J\sum_{\braket{\bm{i}, \bm{j}}}[\hat a_{\bm i}^{\dag}\hat a_{\bm j} + \mathrm{h.c.}] + \frac{U}{2}\sum_{\bm i}\hat n_{\bm i}(\hat n_{\bm i} - 1) + \sum_{\bm i}V_{\bm i}\hat n_{\bm i}.
\end{equation}
Here, $J$ and $U$ designate the tunnelling and on-site interaction energies respectively, $\hat a_{\bm{i}}^{\dag}$ ($\hat a_{\bm{i}}$) is the bosonic creation (annihilation) operator on site $\bm i$, $\hat n_{\bm{i}} = \hat a_{\bm{i}}^\dag \hat a_{\bm{i}}$ is the number operator and $\braket{\bm i, \bm j}$ indicates nearest-neighbour lattice sites.
The effect of the harmonic trapping potential is encoded in the spatial dependence of $V_{\bm i}$ (Methods).
At zero temperature, the homogeneous BHM ($V_{\bm i}=0$) features a quantum phase transition between a superfluid state and a Mott insulating state  \cite{fisher:1989}, occurring for a critical value $\uc$ of the ratio  $u=U/J$.
At finite temperature in the thermodynamic limit, this transition acquires a classical character, {\it i.e.} it is governed by thermal fluctuations in a narrow region around a temperature-dependent critical value   $\uc(T)$.
This thermal transition belongs to the universality class of the 3D classical XY model.
The presence of a harmonic potential $V_{\bm{i}}$ implies that different phases may coexist, and leads to a modification of the critical value (see \figref{fig:fig1}c). 
However, this does not prevent the observation of critical scaling  \cite{donner:2007,zhang:2012, campostrini:2009}. 
In our experiment, the total atom number is $N_{\mathrm{tot}} = \num{4.0(5)e3}$ (Methods), yielding $\braket{\hat n_{\bm i}} \approx \num{1}$ at the centre of the system; the control parameter $u$ is tuned between $u \approx \num{5}$ to $u \approx \num{70}$ by adjusting the optical lattice depth.
\newline

\begin{figure*}[!t]
\centering
\includegraphics[width=\textwidth]{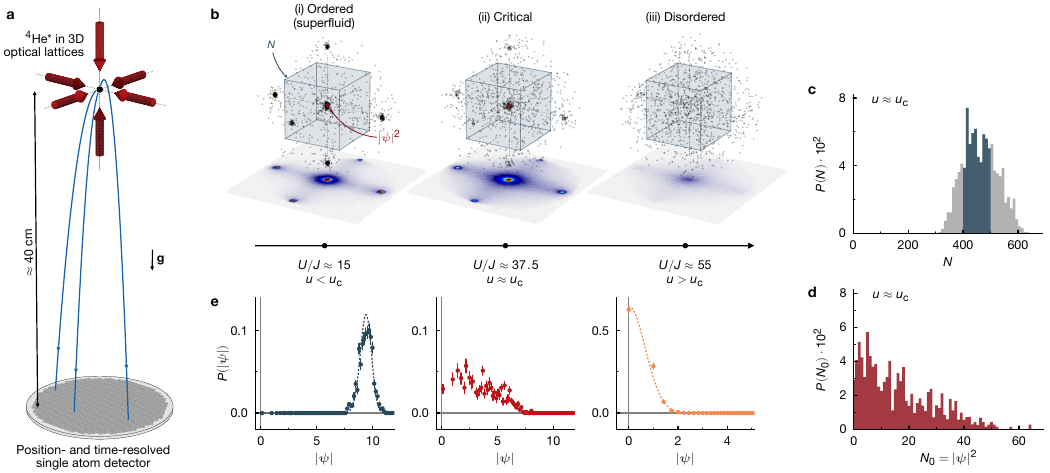}
\caption{
\textbf{Single-atom detection in momentum space and full counting statistics}.
\textbf{a.} Ultracold \he~atoms, released from 3D optical lattices, are detected one-by-one after a free fall onto a micro-channel plate.
Their positions and times of arrival allow us to reconstruct their in-trap momenta.
\textbf{b.} Reconstructed 3D momenta as a function of $u=U/J$.
$|\psi|^2$ and $N$ designate respectively the number of detected atoms in the central mode ($k/\kd \approx 0$, in red), associated with the condensate, and in the first Brillouin zone ($-0.5 \leq k_i/\kd \leq 0.5$, $i = x, y, z$, blue box).
\textbf{c.} Probability distribution of the atom number $N$ near the critical point before (grey) 
and after (blue) post-selection (see Methods).
Post-selecting on $N$ reduces the effect of shot-to-shot fluctuations on the statistics of $N_0$ (SI, section~\ref{SM:postselection}).
\textbf{d.} Probability distribution of the condensate atom number $N_0 = |\psi|^2$ near the critical point,  obtained from about 1200 experimental repetitions.
\textbf{e.} Probability distributions of  the order parameter $|\psi|$.
The dashed lines in (i) and (iii) correspond to the expected shapes for the distributions calculated from the measured mean value $\braket{|\psi|^2}$ (see text).
In this figure and all other figures, error bars represent statistical uncertainties obtained from a bootstrap analysis (see Methods).
}
\label{fig:fig2}
\end{figure*}

In classical field theory, the 3D XY transition is associated with a complex-valued order parameter $\psi = |\psi|\ee^{\ii\phi}$. In our quantum system, the order parameter is a field operator $\hat{\psi}$ destroying a particle in the condensate mode \cite{pitaevskii:2016}. Its squared amplitude  $\hat N_0 = \hat \psi^\dag\hat\psi$ is an observable that corresponds to the integer-valued number of particles $N_0$ in the $\bm{q} = \bm{0}$ quasi-momentum mode.
The ordered phase manifests itself through a macroscopic population $N_0 \gg 1$. Our detection method --- specific to metastable helium  \cite{vassen:2012} --- is ideally suited to reconstruct the full statistics of $|\psi|$ from measuring $|\psi|= \sqrt{N_0}$ in each experimental shot, as it probes the 3D momentum distribution atom-by-atom (\figref{fig:fig2}a,b).
More details on the  apparatus can be found in our previous works \cite{cayla:2018a, tenart:2021a, bureik:2025} and in the Methods.
We show in \figref{fig:fig2}b exemplary single-shot experimental realisations.
For $u < \uc$, we observe the 3D diffraction pattern associated with a superfluid with long-range phase coherence.
Deep in the normal (finite-temperature Mott insulator) regime ($u > \uc$), phase coherence is lost and the  distribution is featureless. 
For each shot we denote by  $N$ the total detected atom number in the first Brillouin zone (blue cube in \figref{fig:fig2}b)
and $N_0=|\psi|^{2}$ the detected atom number in a sphere of radius $|\delta \bm{k}| = \num{0.03}\kd$ centred on $\bm{k} = \bm{0}$ (Methods).
Here, $\kd = 2\pi/d$ is the lattice momentum, with $d$ the lattice constant.
The statistics of $N_0$ are slightly influenced by the fluctuations of  $N$ (see Supplementary Information (SI), section \ref{SM:postselection}). 
Post-selecting on $N$ (\figref{fig:fig2}c) allows us to evaluate and reduce this effect.

We reconstruct the probability distribution function $P(|\psi|)$  from the full counting statistics of $N_0$ (see Methods).
The PDFs shown in \figref{fig:fig2}e exhibit drastic differences with $u$.
In the ordered (superfluid) regime, the distribution matches a Gaussian with  a maximum around 
$|\psi_0 | \approx 10$, capturing the macroscopic population of the condensate  and its expected statistics \cite{herce:2023}.
In the disordered (Mott) phase, the PDF is different, taking a maximum value for $|\psi_0| = 0$.
It is well fitted by a (truncated) Gaussian distribution, as expected from thermal statistics \cite{carcy:2019a, herce:2023}.
The PDF near the phase transition $u=37.5$ strongly contrasts with both the ordered and disordered regimes.
It features a relatively large width capturing the large fluctuations of the order parameter near $u_{c}$ and a shape  distinct from Gaussian.
\newline

\begin{figure*}[!t]
\centering
\includegraphics{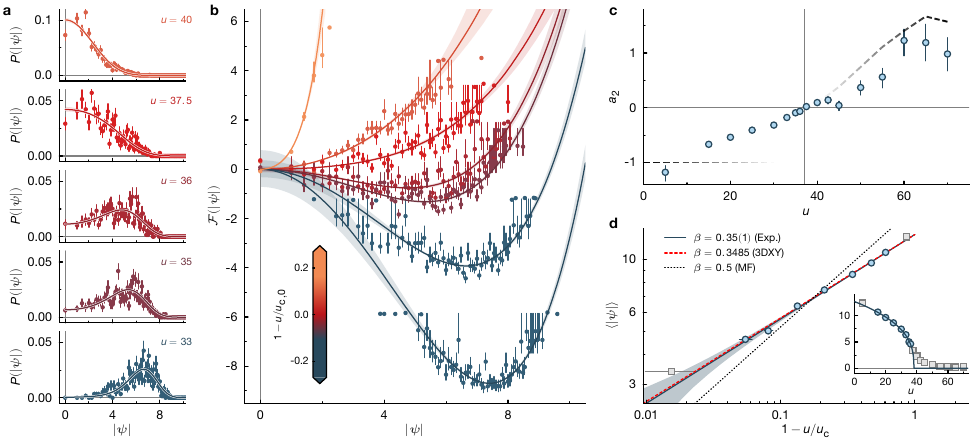}
\caption{
\textbf{Non-Gaussian probability distribution of the order parameter.}
\textbf{a.} Probability distribution function of the  order-parameter amplitude $|\psi|$ near the transition in the range $33 \leq u \leq 40$.
The solid lines are fits of the form $P(|\psi|) = \ee^{- \mathcal{F}(|\psi|)}$, where $\mathcal{F}(|\psi|)$ is given by Eq.~\eqref{eq:Ffunc}.
\textbf{b.} Effective thermodynamic potential $\mathcal{F}(|\psi|)$ reconstructed from the experimental data depicted in panel (\textbf{a}), together with $u = 30$ and $u = 55$.
\textbf{c.} Fitted coefficient $a_2$ (see Eq.~\eqref{eq:Ffunc}) as a function of $u$.
The dashed lines correspond to the expected asymptotic behaviours, for $u \ll \uz$ and $u \gg \uz$.
In the former case, the Poisson statistics of $|\psi|^{2}$ yield $a_2 = -1$, while in the latter case, where statistics are thermal, one finds $a_2 = -\ln [\braket{N_0} / (1 + \braket{N_0})]$.
The sign change of $a_2$ marks the critical point $\uz$ with $\uz = \num{37.1(2)}$.
\textbf{d.} Average value $\braket{|\psi|}$ as function of $\uc-u$ 
for $u<\uc$, showing power-law  scaling  $\braket{|\psi|} \propto (\uc-u)^{\beta}$.
The fit gives $\uc = \num{38.1(2)}$ and a 
critical exponent $\beta=\num{0.35(1)}$ compatible with $\beta_{\rm 3D XY}=0.3485$ (red dashed line), but incompatible with the mean field prediction $\beta_{\rm MF}=0.5$ (black dotted line).
The inset shows the same quantity as a function of $u$, with the blue dots representing experimental data points used to extract the coefficient $\beta$.
}
\label{fig:fig3}
\end{figure*}

The shape of the order-parameter PDF  is captured by introducing an effective thermodynamic potential $\mathcal F(|\psi|)=-\ln P(|\psi|)$.
Close enough to the transition, it is well approximated by a low-order polynomial
\begin{equation}
    \mathcal F(|\psi|) \simeq a_0 + a_2 |\psi|^2+a_4|\psi|^4,
\label{eq:Ffunc}
\end{equation}
similar in spirit to Landau's approach to phase transitions.
As detailed in the SI, this thermodynamic potential efficiently describes the statistics of the order-parameter amplitude, regardless of the universality class.
It has a non-trivial minimum for $a_2< 0$ (ordered phase) and a minimum at $|\psi|=0$ for $a_2>0$ (disordered phase).
Near the transition, $P(|\psi|)$ becomes highly non-Gaussian when $a_2^2/a_4\lesssim 1$.

In \figref{fig:fig3}a, we plot  $P(|\psi|)$  measured near the transition, while \figref{fig:fig3}b shows the corresponding potentials.
We observe a continuous change from a distribution similar to that
of the ordered phase ($u=30$),  to a distribution closer to the disordered phase ($u=55$),
with flatter distributions in between.
Equation~\eqref{eq:Ffunc} fits well our experimental data (solid lines in \figref{fig:fig3}a,b), with the fitted coefficient $a_2$ shown in \figref{fig:fig3}c.
As anticipated, $a_2$ increases from negative to positive values.
Empirically, we identify a critical value $\uz = \num{37.1(2)}$ where $a_{2}$ is equal to zero.
We stress that, away from the thermodynamic limit, $\uz$ is different from $\uc$, the critical point associated with universal scaling.
This result is well-established in the phenomenology of finite-size systems \cite{balog:2022} (see SI, section 5). 
 
It might seem surprising that our measurements align with the Landau-like mean-field theory as the probed transition belongs to the 3D XY universality class. 
However, at fixed $u$, beyond-mean-field effects  manifest themselves only in the tails of the order-parameter PDF, and correspond to rare events \cite{balog:2025} (see SI, section \ref{SM:PDF}).
Experimentally, the typical number ($\sim 700$) of shots used to construct a PDF, although large, remains insufficient to resolve these rare events. 
In contrast, beyond mean-field effects are revealed by measuring the PDFs upon varying the control parameter. 
This is shown by our measurement of the average value $\braket{|\psi|}$ as a function of $u$
(\figref{fig:fig3}d).
First, $\braket{|\psi|}$ exhibits power-law scaling with a critical exponent $\beta=0.35(1)$ compatible with that of the 3D XY universality class, $\beta_{\rm 3D XY}=0.3485$ \cite{campostrini:2001}, and distinct from the mean-field prediction $\beta_\mathrm{MF}=0.5$ (\figref{fig:fig3}d).
Second, the critical scaling for $\braket{|\psi|}$ is found even for small values of $u$, suggesting that the critical regime extends deeply into the ordered (superfluid) phase. 

The power-law  scaling of $\braket{|\psi|}$ expected in the thermodynamic limit is observed where the order-parameter PDFs are well captured by Gaussian statistics {\it i.e.} where $\xi < L$ as illustrated in \figref{fig:fig1}a.
Conversely, non-Gaussian statistics appear in a narrower region close to the critical point where finite-size effects become significant and lead to a breakdown of power-law  behaviour.
Moreover, the harmonic trap leads to sizeable corrections to $\uc$, as compared to predictions for a homogeneous system (see \figref{fig:fig1}c).
The measured value of $\uc = \num{38.1(2)}$ agrees well with predictions based on Quantum Monte Carlo (QMC) calculations in a harmonic trap \cite{herce:2021}. 
The fact that $\uc$ exceeds the critical value ($u\approx 29.3$) for the superfluid-Mott-insulator quantum phase transition with filling $n=1$ highlights the fact that the transition occurs in the outer region of the trap, where $n<1$ (see Methods and \figref{fig:fig5}).
\newline

To quantify explicitly the non-Gaussian nature of the order-parameter statistics close to the critical point, we now focus on its high-order cumulants.
Cumulants at all orders reconstruct the full statistical information contained in the PDF, similarly to moments. 
The cumulant methodology offers a dual advantage.
First, cumulants are a standard tool in statistics for capturing non-Gaussian behaviour;
the presence of any cumulant $\kappa_{n \geq 3} \neq 0$ is necessary and sufficient for non-Gaussian statistics.
Second, as we shall see, the dependence of cumulants on the control parameter proves to be extremely rich, providing new and surprising insights into the critical behaviour. 

\begin{figure}[!t]
\centering
\includegraphics[scale=1]{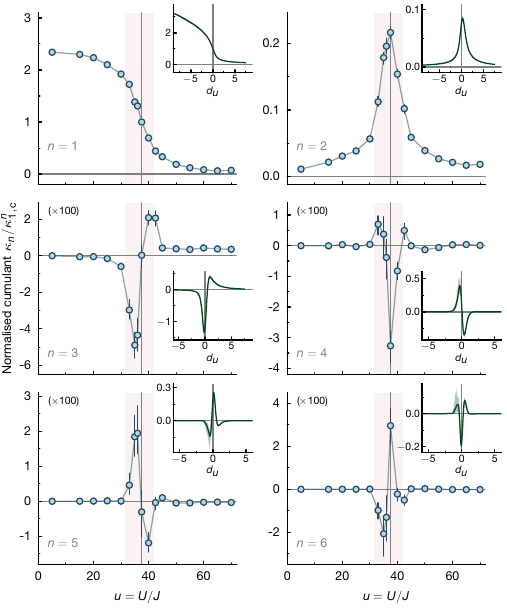}
\caption{
\textbf{Cumulants of the order parameter across the transition}.
Normalised cumulants $\kappa_n/\kappa_{1,c}^{n}$ of the order parameter $|\psi|$ up to order $n=6$ as a function of $u = U/J$, with $\kappa_{1,c} \equiv \kappa_{1}(u=\uz)$ (see SI). 
Significant deviations from zero, observed near the transition point $u = \uz$ for cumulants of order $n \geq 3$, signal the presence of non-Gaussian fluctuations.
The inset shows the universal curve for each cumulant, calculated from a homogeneous quantum rotor model, belonging to the 3D XY universality class.
The control parameter is rescaled according to $d_u = A(T)(u/\uc - 1)L^{1/\nu}$, where $\nu = 0.6718$ is a critical exponent, $L$ the linear system size and $A(T)$ a temperature-dependent coefficient (see text and SI).
}
\label{fig:fig4}
\end{figure}

\begin{figure}[!t]
\centering \includegraphics[width=\columnwidth]{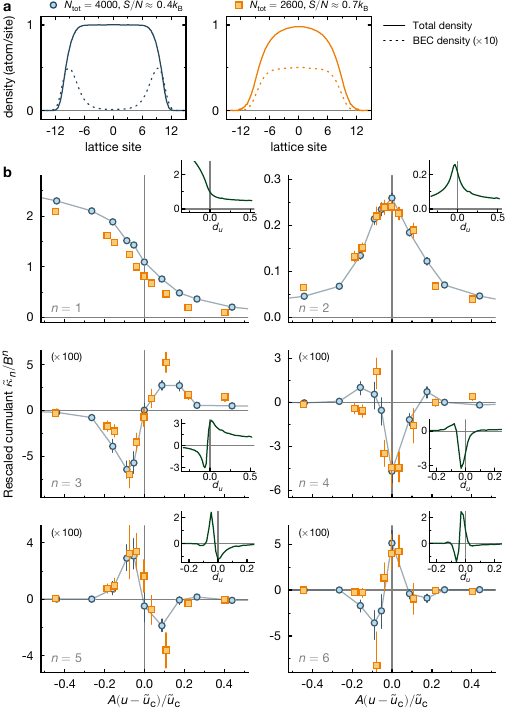}
\caption{
\textbf{ Robustness of the cumulant critical behaviour}
\textbf{a.} In-trap condensate density (dashed line) and total density (plain line), calculated using QMC under the experimental conditions associated with two experimental data, near their respective critical point $u \approx \uz$. 
The volume $V$ of the simulation box is $V = 30^3$ lattice sites. 
\textbf{b.} Measured cumulants expressed in reduced units for the comparison of the two datasets (blue dots correspond to the data set shown in \figref{fig:fig4}). 
The amplitude $\tilde{\kappa}_n=\kappa_n/B^n$ is plotted as a function of a dimensionless distance $A (u-\tilde{u}_{\mathrm{c}})/\tilde{u}_{\mathrm{c}}$, 
showing  that cumulants coincide at all orders $n$.
$A$, $B$ and $\tilde{u}_{\mathrm{c}}$ are scale factors determined through a collapse optimisation of both datasets (see Methods).
The insets show cumulants obtained from numerical simulations of the quantum rotor model with $L=10$ and $T/2Jn_{\mathrm{QR}} = 0.05$, \emph{i.e.} in a regime of substantial quantum corrections to the universal scaling (SI).\\
}
 \label{fig:fig5}
\end{figure}

The measured cumulants $\kappa_n(|\psi|)$ of $|\psi|$ are shown in \figref{fig:fig4}.
Cumulants $\kappa_1 = \braket{|\psi|}$ and $\kappa_2 = \braket{|\psi|^2} - \braket{|\psi|}^2$ behave as expected from the phenomenology of a continuous phase transition; $\kappa_1$ is finite in the ordered phase ($u \ll \uc$) and vanishes in the disordered phase ($u \gg \uc$), while $\kappa_2$, which quantifies the order-parameter fluctuations and can be associated with a susceptibility \cite{nishimori:2010}, exhibits a peak at $u \approx \uc$.

The evolution of higher-order cumulants with  $u$ is instead more instructive. 
First, cumulants of order $n \geq 3$ vanish deep in the ordered ($u \ll \uc$) and disordered ($u \gg \uc$) phases, while they take large values near the phase transition $u\approx \uc$.
These variations reveal the crossover of the order-parameter statistics from Gaussian to non-Gaussian  across a continuous phase transition~\cite{Note1}.
A similar phenomenology has been  proposed to identify the phase transition from a hadron gas to a quark-gluon plasma in quantum chromodynamics (QCD) \cite{stephanov:2011, asakawa:2009, bzdak:2020}. 
Second, we observe abrupt sign changes of the high-order cumulants  near the phase transition.
We discuss this unexpected observation below.
Before proceeding, we note that the harmonic confinement present  in our experiment leads to the coexistence of locally superfluid (or critical) and normal regions. 
Yet the observed non-Gaussian fluctuations do not stem from the phase coexistence making the distribution \emph{e.g.}, bimodal (bimodality is not observed). 
This is because we probe the statistics of a \emph{global}, rather than \emph{local}, order parameter, which is always dominated by the superfluid or  critical regions in the trap whenever they exist. 

In \figref{fig:fig4}, we first compare the measured cumulants with those obtained numerically in homogeneous models belonging to the  3D XY universality class and find similar sign changes of the cumulants (insets).
A finite-size scaling analysis performed on these models --- defined on homogeneous lattices with periodic boundary conditions --- shows an excellent collapse (see \figref{fig:figS1} in the SI).
This demonstrates that the order-parameter cumulants exhibits universal scaling functions across the phase transition in homogeneous  systems.

While both experiment and simulations display sign changes, we observe qualitative differences for cumulants of order $n\geq4$. 
In the critical regime ($L,\xi\gg \ell_G$), universal scaling functions depend not only on the universality class but also on the geometry of the system (boundary conditions, shape of the system and presence of a trap) \cite{vasilyev:2009, hucht:2011}. However, these factors alone cannot account for the differences observed between simulations and experiment in Fig.~\ref{fig:fig4}, as we shall now explain. First, numerical simulations of a non-homogeneous classical model yield the same qualitative results as those of homogeneous models (see SI, \figref{fig:figS11}).
Second, we experimentally tested the effect of the system geometry and found that it does not affect our observations. 
To this aim, we probed the superfluid transition in a different microscopic configuration 
(see Methods and \edfigref{fig:ed1}).
There, the critical behaviour develops at the trap centre (leading to a substantial shift of $\uc$), in contrast with the first dataset in which the transition occurs in the outer region of the trap.
The results in \figref{fig:fig5} show similar behaviour of the cumulants of $|\psi|$ in these two microscopic configurations, confirming that the observed deviations from the classical results are robust to changes in the system geometry. Remarkably, the rescaled units in \figref{fig:fig5} make all cumulants $n\geq2$ coincide, even though they are independent of one another.

Instead, we attribute the differences between experiment and 3D XY predictions to universal corrections to scaling induced by quantum fluctuations. Indeed, similar corrections are observed in a model of quantum rotors at low temperatures (see insets of \figref{fig:fig5}b and SI, \figref{fig:figS10}). Only higher-order cumulants are sensitive to these corrections to scaling that affect mostly the tails of the PDF \cite{balog:2025}. 
Our observations highlight that cumulants provide high resolution in the reconstruction of the critical behaviour, unveiling subtle quantum effects at thermal transitions.
\newline

In conclusion, we measured the  statistics of the order-parameter amplitude across a continuous phase transition using ultracold lattice bosons.
Our experiment is performed in a harmonic trap and realises an inhomogeneous system of mesoscopic size.
Interestingly, we observe many of the emblematic predictions for homogeneous systems, including the change in the effective thermodynamic potential --- similar in spirit to Landau theory --- and the non-Gaussian critical regime. 

In addition, we experimentally unveiled abrupt sign changes of the order-parameter cumulants across the transition. We provide numerical evidence and experimental indications of their universal character.
These sign changes are general features of criticality, which are not specific to the universality class explored here.
They have been observed numerically in the $O(N)$ universality classes \cite{pan:2013} and stem from the shape of the critical PDF across the transition \cite{rancon:2025}.
Furthermore, they are considered as promising observables to identify phase transitions in the QCD phase diagram \cite{stephanov:2011, asakawa:2009, bzdak:2020}.

The standard approach of probing critical scaling is achievable in macroscopic systems with negligible finite-size effects.
Conversely, our measurement of the order-parameter statistics demonstrates a complementary approach that is well suited to mesoscopic systems. 
This approach is highly sensitive to the nature of critical fluctuations, including the presence of a quantum phase transition at $T=0$, and promising for several quantum-simulation platforms
\cite{xu:2020a,jurcevic:2017, makhalov:2019, ebadi:2021, chen:2023, ho:2025}.
Furthermore, the experimental ability to reconstruct the order-parameter fluctuations in finite-size inhomogeneous systems challenges theoretical and numerical methods for quantum systems.


\section*{Methods}

{\it Production of low-entropy and low-number BECs.}
Our experiment uses standard laser cooling and evaporative cooling steps to produce a spin-polarized Bose-Einstein condensate (BEC) of $\sim \num{4e5}$ metastable helium-4 (\he) in the sub-level $m_J=1$ of the state $2^3$S$_1$ \cite{bouton:2015}.
The atom number is then reduced to $\num{4.0(5)e3}$ ($\num{2.6(3)e3}$ for the second dataset) using state-dependent inelastic losses: we use two-photon Raman transitions to transfer a controlled fraction of the atoms to the sub-level $m_J=0$ where two-body Penning ionization leads to losses.
The small BEC is then loaded adiabatically in the cubic optical lattice. 
In the 3D lattice, the trapping frequency of the harmonic potential $V_{\bm{i}}$ is $\omega /2 \pi = 140(5) \sqrt{V_0}~$Hz, with $V_0$ the lattice amplitude in unit of the recoil energy $E_{\rm R}=h^2 /8m d^2$, and $m$ the mass of an atom.
A thermometry based on the comparison with ab initio QMC calculations \cite{carcy:2021} yields a temperature in the range $k_{\rm B}T/J \sim 1.5-2$.
This is compatible with an entropy per atom of $S/N \sim \num{0.4(1)} k_{\rm B}$ ($\num{0.7(1)}k_{\rm{B}}$ for the second dataset) extracted from the measured condensate fraction before loading the BEC in the optical lattices \cite{carcy:2021}.
\\

{\it Volume used to compute the statistics of $|\psi|$.}
In translationally invariant lattice systems where the quasi-momentum ${\bm q}$ is a good quantum number, the number of condensate atoms $N_0$ is equal to the atom number in the mode ${\bm q}={\bm 0}$.
In a finite-size and harmonically-trapped gas, $N_0$ remains well approximated by the atom number at ${\bm q}={\bm 0}$.
In this work, we probe the gas in the momentum basis where the modes located at $\bm n \kd$, with $\bm n \in \mathbb{Z}^{3}$, are copies of the mode ${\bm q}={\bm 0}$.
We use a sphere of radius $0.03 \kd$ centred on the central peak (${\bm n}={\bm 0}$) or first-order diffraction peaks  ($\| \bm n \|=1$) to measure $N_0$ in each shot.
This radius is unambiguously determined from the width of the bosonic bunching signal which is set by the size of one mode \cite{carcy:2019a, cayla:2020}.
As shown in the SI, the exact value of the radius we use in the analysis does not affect our observations.
\\

{\it Probability Distribution Function (PDF) and cumulants of $|\psi|$.}
The PDF of $|\psi|=\sqrt{N_0}$ is reconstructed from the histogram of the measured values of the condensate atom number $N_0$ in each shot. The histogram bins for the discrete variable $N_0$ are the positive integers and the histogram is normalised such that it sums to one, $\sum_n P(N_0=n) = 1$.
The PDF of $|\psi|$ is obtained from remarking that $P(N_0=n) = P(|\psi|=\sqrt{n})$. Its normalisation is set by that of the histogram on $N_0$.\\
The cumulants of $|\psi|$ in Fig.~\ref{fig:fig4} and Fig.~\ref{fig:fig5}b are computed directly from the measured values of $\sqrt{N_0}$. They are normalised by $\langle \sqrt{N} \rangle^n$ to account for the variation of the atom number $N$ in the first Brillouin zone with the interaction parameter $u$.
\\

{\it Bootstrap procedure.}
The errorbars on experimental cumulants, probability distribution functions and fit parameters are $68 \%$ confidence intervals obtained from a bootstrap method.
500 pseudosamples are generated from sampling the experimental shots with replacement.
The observables are computed for each pseudosample.
The resulting histograms take a typical bell-shape curve, from which the mean, the $16^\mathrm{th}$ and $84^\mathrm{th}$ percentiles are extracted. The percentiles provide an estimate of the statistical uncertainty on the mean.
\newline

\begingroup
\renewcommand{\thefigure}{1}
\renewcommand{\theHfigure}{ED1}
\renewcommand{\figurename}{Extended Data Fig.}
\begin{figure}[!t]
\begin{center}
\includegraphics[scale=1]{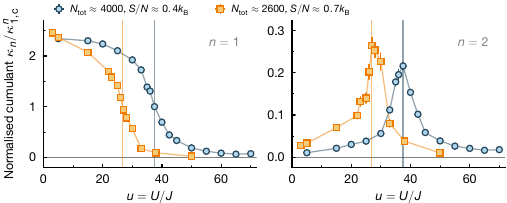}
\caption{
\textbf{Mean and variance of the order parameter in two distinct microscopic configurations.}
Normalised mean $\kappa_1/\kappa_{1,c}$ and variance $\kappa_2/\kappa_{1,c}^2$ of the order-parameter amplitude measured in the two datasets analysed in \figref{fig:fig5}.
The vertical lines depict the critical values $\uz$ of the respective datasets}
\label{fig:ed1}
\end{center}
\end{figure}
\endgroup

{\it Trapped atomic densities at different total atom number.}
In \figref{fig:fig5}, we compare the order-parameter cumulants measured in lattice gases with different total atom numbers and entropy per atom.
As discussed in the main text and shown in \edfigref{fig:ed1}, the difference in the measured critical values $\uz$ highlights the distinct configurations of in-trap densities close to $\uz$: at small atom number the condensate develops at the trap centre; in contrast the condensate appears in the outer region of the trap at large atom number.
These differences are confirmed by numerical calculations of the harmonically trapped Bose-Hubbard Hamiltonian using QMC \cite{carcy:2021}, see \figref{fig:fig5}a.
There we contrast the local density $n_i = \braket{\hat a_i^\dagger \hat a_i}$ with the condensate density $n^{(0)}_i = \frac{1}{V} \sum_{j} \braket{ \hat a_i^\dagger \hat a_j }$, where $V$ is the volume of the simulation box. All the parameters used in these numerical calculations are calibrated from the experiment, except for the temperature, which is estimated by matching the momentum distribution calculated numerically via QMC with the one measured experimentally, as described in \cite{carcy:2021}.
\\

{\it Cumulants in rescaled units.}
The cumulants measured for different microscopic configurations of the trapped gas are shown in \figref{fig:fig5} in rescaled units inspired by theoretical approaches to critical collapse. 
The ratio $u=U/J$ is replaced by the distance $A \delta u$ to the critical point, with $\delta u = (u-\tilde{u}_{\mathrm{c}})/\tilde{u}_{\mathrm{c}}$, and the amplitudes of the cumulants $\kappa_n$ are rescaled as $\tilde{\kappa}_n = \kappa_n / B^n$. 
The parameters $A$, $B$ and $\tilde{u}_{\mathrm{c}}$ are specific to a microscopic configuration while the function $\tilde{\kappa}_n(A \delta u)$ is expected to be independent of the configuration when it describes universal behaviour.

In our procedure, the non-universal scaling factors are obtained by minimising the cumulative error (the sum of squared differences computed over the second through sixth cumulants) between the two datasets after rescaling.
We arbitrarily chose the second dataset (orange squares in \figref{fig:fig5}) as a reference --- setting $A=1$ and $B=1$ and the value $\tilde{u}_{\mathrm{c}}=\uz$ determined from $a_2=0$ --- 
and numerically optimise the scaling factors $A$, $B$ and $\tilde{u}_{\mathrm{c}}$ of the first dataset (blue circles in Figs.~\ref{fig:fig4}, \ref{fig:fig5}).
Note that, because the data points are discrete and not aligned after rescaling, the values of $\tilde\kappa_n$ for one dataset are interpolated linearly to compute the error with respect to the other dataset.
The fitted values for the first dataset are $A=\num{1.32(8)}$, $B = \num{0.91(1)}$ and $\tilde{u}_{\mathrm{c}} = \num{37.5(1)}$, \emph{i.e.} the fitted value for $\tilde{u}_{\mathrm{c}}$ is close to $\uz = \num{37.1(2)}$ determined from $a_2=0$ (see main text).
\newline

\textbf{Acknowledgements ---}
We thank I. Balog, A. Browaeys, Z. Hadzibabic and the members of the Quantum Gas group at Institut d'Optique for insightful discussions. 
Numerical simulations were conducted on the CPBsmn cluster at the ENS of Lyon.
\newline

\textbf{Funding ---}
We acknowledge funding from the R\'egion Ile-de-France in the framework of the DIM QuanTiP, the ``Fondation d'entreprise iXcore pour la Recherche'', the French National Research Agency (Grant number ANR-17-CE30-0020-01) and France 2030 programs of the French National Research Agency (Grant number ANR-22-PETQ-0004). A.R. has benefited from the financial support of the Grant No. ANR-24-CE30-6695 FUSIoN.
\newline


\clearpage
\newpage
\makeatletter
\renewcommand{\thefigure}{S\arabic{figure}}
\renewcommand{\theequation}{S\arabic{equation}}

\renewcommand{\theHfigure}{S\arabic{figure}}
\renewcommand{\theHequation}{S\arabic{equation}}
\makeatother
\setcounter{figure}{0}
\setcounter{table}{0}
\setcounter{section}{0}
\setcounter{equation}{0}
\appendix

\makeatletter
\let\oldtitle\@title
\title{Supplementary Information for: \\ \oldtitle}
\makeatother

\maketitle

\subsection{Effect of the total atom number on the order-parameter statistics}
\label{SM:postselection}

The statistics of the order parameter could in principle be affected by technical noise, \emph{i.e.}, by the shot-to-shot fluctuations of the total atom number.
To mitigate this contribution we post-select experimental shots on the total atom number $N=N_{\rm FBZ}$ in the first Brillouin zone. Our data-taking procedure consists of measuring at different values of $u$ in consecutive shots, so that the atom number statistics is guaranteed to be independent of $u$.
For instance, we observe in the experiment the same variation of $N$ with $u$ as obtained in Quantum Monte Carlo simulations of the Bose-Hubbard Hamiltonian in the presence of an harmonic trap.
These variations simply reflect the change in the shape of Wannier functions with $u$, which in turn affects the number of atoms measured in the first Brillouin zone. 
Note that $N$ is slightly under-estimated in the superfluid regime as the central peak saturates the detector due to its high atomic density.
The scaling analysis of \figref{fig:fig3}d, in particular, is performed on the diffracted peaks at $\bm k = \bm \kd$ to avoid the saturation effects.
We discuss this in more detail in section \ref{sec:momentum}.

\begin{figure}[!t]
\centering
\includegraphics{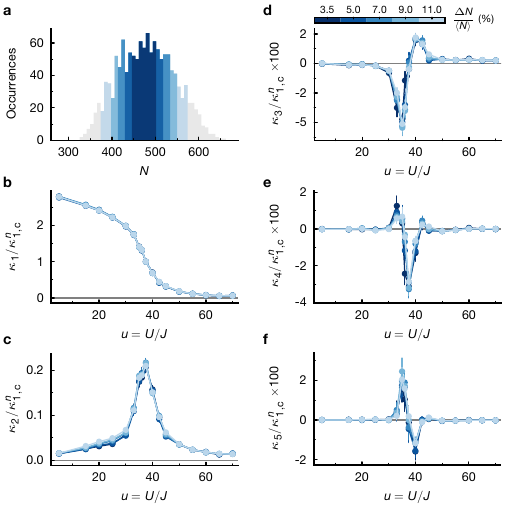}
\caption{
\textbf{Post-selection on total atom number $N$.}
\textbf{a.} Distribution of total atom numbers, similar to what is shown in \figref{fig:fig2}d.
The analysis relies on post-selection around an average value $\braket{N}$ of RMS width $\Delta N$.
\textbf{b-f.} Cumulants calculated for several $\Delta N$.
The quantitative behaviour of the cumulants remain unchanged when varying the post-selection width.
}
\label{fig:figS3}
\end{figure}

\figref{fig:figS3} presents the cumulants of the order parameter obtained when varying the amplitude $\Delta N$ of the fluctuations around a fixed mean value $\braket{N}$ in the post-selection on the total atom number. This illustrates that our results are only marginally affected by the fluctuation amplitude $\Delta N$.

\subsection{Effect of the measurement volume on the order-parameter statistics}

\begin{figure}[!t]
\centering
\includegraphics{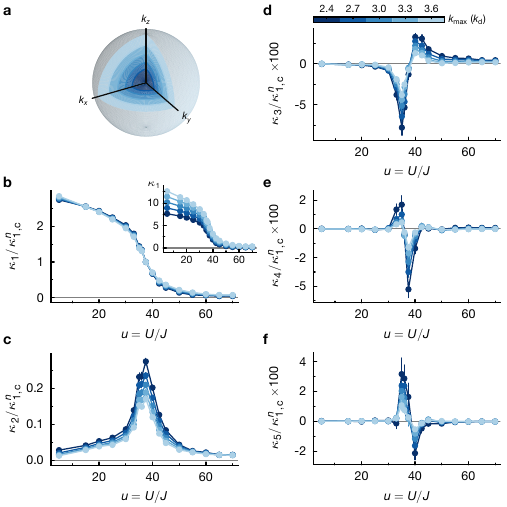}
\caption{
\textbf{Cumulants of the order parameter vs measurement volume.}
The volume defining $N_0$ is varied from $0.024\ k_d$ to $0.036\ k_d$.
The qualitative behaviour of the cumulants remains unchanged when slightly varying the measurement volume around the chosen value $0.03\ k_d$.
}
\label{fig:figS4}
\end{figure}

For each experimental shot, we count the atom number $N_0$ in the condensate mode (from which we deduce $|\psi|$), in a spherical volume centred on $\bm{k}=\bm{0}$. We choose the radius of this sphere to match the size of a mode in momentum space, as obtained from the width of the bunching peak \cite{tenart:2021a}. We show in \figref{fig:figS4} that small variations of the radius around this value does not modify the measured cumulants significantly. The small variations are instead expected: the increase of $\kappa_1$  with volume reflects the increase in the atom number $N_0$; the contrast of higher-order cumulants decreases with  volume as other modes are mixed with the condensate mode.

\subsection{Momentum vs quasi-momentum}
\label{sec:momentum}

The condensate order parameter $|\psi| \sim \sqrt{ N_{\bm{q} = \bm{0}} }$ is associated with the mode with quasi-momentum $\bm q = \bm 0$.
Time-of-flight measurements as performed in this work give instead access to populations in momentum modes, the in-trap quasi-momenta being projected onto the plane wave basis as the lattice is abruptly switched off.
The quasi-momentum mode $\bm{q} = \bm{0}$ is replicated at the reciprocal lattice vectors $\kd\bm{n}$, with $\kd = 2\pi/d$ ($d = \SI{775}{\nm}$ is the lattice constant) and $\bm n \in \mathbb{Z}^{3}$.
This mechanism is analogous to a beam-splitter with multiple outputs and leads to a multinomial noise of each of the output modes $\bm k = \bm n \kd$.

\begin{figure}[!t]
\centering
\includegraphics[scale=1]{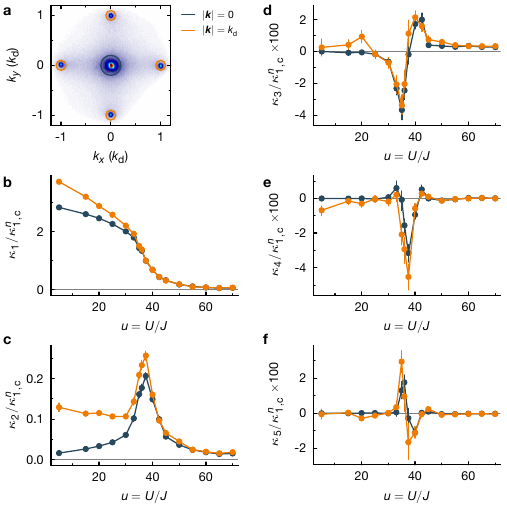}
\caption{
\textbf{Cumulants on diffraction peaks.}
Comparison of the cumulants calculated from the mode $\bm k=\bm 0$ (blue dots) and the first-order diffraction modes $\bm k = \bm n k_d$ with $|\bm n |=1$ (orange dots).
}
\label{fig:figS5}
\end{figure}

In addition, the average populations of the momentum peaks $\bm k = \bm n \kd$ are not all proportional to that of the quasi-momentum mode $\bm q = \bm 0$.
This effect results from the variation of the Wannier functions with $u$, whose envelop in momentum space modifies the amplitudes of the diffracted peaks.
Numerically, we find that the average population of the first-order peaks $|\bm n| = 1$ is proportional to that of the quasi-momentum mode $\bm q = \bm 0$, independently of $u$.
In other words, the ratio $\braket{N_{\bm{k} = \pm \bm{k}_\mathrm{d}}}/\braket{N_{\bm{q} = \bm{0}}}$ is constant. 
In contrast, the ratio $\braket{N_{\bm{k} = \bm{0}}}/\braket{N_{\bm{q} = \bm{0}}}$ varies with $u$ similarly to $\braket{N}$. 
As a result, one must consider the normalised variable $N_{\bm{k} = \bm{0}}/\braket{N}$ to obtain a variation with $u$ that faithfully reflects that expected for the variable $N_{\bm{q} = \bm{0}}$. 
To verify this assertion, we compare the cumulants of $\sqrt{N_{\bm{k} = \bm{0}}}/\sqrt{\braket{N}}$ with those of $\sqrt{N_{\bm{k} = \pm \bm{k}_\mathrm{d}}}$ in \figref{fig:figS5}.

In \figref{fig:figS5}, we plot the cumulants normalised to the value of their average $\kappa_{1,\rm c}$ measured at $u=\uz$.
The high-order cumulants ($n\geq3$) are almost identical. The first and second order cumulants slightly differ instead, with differences which are well understood.
The lower values of $\kappa_1/\kappa_{1,\rm c}$ of the central peak $\bm{k}=\bm{0}$ are  due to a saturation of the Micro-Channel Plate detector occurring at small values of $u$ where peak densities are the largest. The second order cumulants $\kappa_2/\kappa_{1,\rm c}^2$ differ mainly due to stronger effect of the multinomial process for the diffracted peaks at $\pm \bm{k}_{\rm d}$, with respect to that occurring for the peak $\bm{k}=\bm{0}$. 

The analysis of the cumulants of the central and first-order peaks shows how reliable the observation of oscillating high-order cumulants near the transition point is. Furthermore, it validates the study of the cumulants of the variable $\sqrt{N_{\bm{k} = \bm{0}}}/\sqrt{\braket{N_{\rm FBZ}}}$ in order to quantitatively reflect the properties of the quasi-momentum mode $\bm{q}=\bm{0}$. All the cumulants shown in the main text and in the SI for the central peak $\bm{k}=\bm{0}$ refer to this variable.

\subsection{Critical behaviour of $\braket{|\psi|}$} 

\begin{figure}[!t]
\begin{center}
\includegraphics[scale=1]{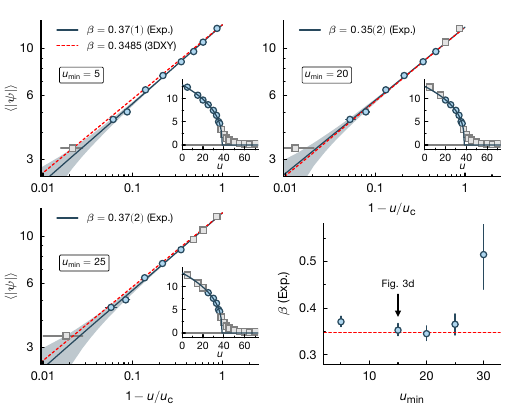}
\caption{
\textbf{Universal scaling of $\braket{|\psi|}$.}
Scaling analysis similar to that of \figref{fig:fig3}d, for different ranges $u_{\rm min} \leq u \leq 37$.
The bottom right plot shows the extracted critical exponent $\beta$ as a function of $u_{\rm min}$. The dashed red line corresponds to the expected value of the critical exponent $\beta$ for the 3D XY universality class.
}
\label{fig:figS7}
\end{center}
\end{figure}

Here we complement the analysis of the critical scaling of the order-parameter amplitude, shown in \figref{fig:fig3}d, by fitting the data over different ranges of $u$ values. In particular, we fit the data over windows $u_{\rm min} \leq u \leq 37$ with a varying minimum value $u_{\rm min}$. The results are illustrated in \figref{fig:figS7}.

For $u_{\rm min} = 5$, corresponding to using all the available data with $u \leq 37$, we observe a small discrepancy between the experimental value of $\beta$ and the expected value in the 3D XY universality class.
For $u_{\rm min} = 15, 20$ or $25$, the measured and 3D XY critical exponents are compatible, suggesting that the critical regime, described by scaling and universality, extends at least down to $u = 15$.
For a larger value $u_{\rm min} = 30$, the measured exponent $\beta$ disagrees significantly with the 3D XY universality class, although with much larger errorbars.
This discrepancy occurs in the range of $u$ where high-order cumulants become non-zero, signalling non-Gaussian statistics associated with finite-size effects ($\xi$ is of the order of $L$). It could therefore originate from the range of $u$ values becoming too narrow around $u_c$ to observe the correct critical behaviour emerging in the thermodynamic limit.

\subsection{Fitting of the probability distribution functions}

\begin{figure}[!t]
\centering
\includegraphics{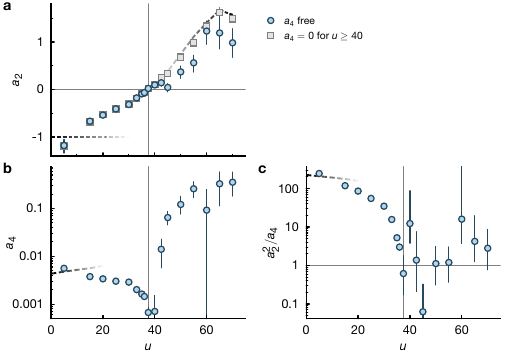}
\caption{
\textbf{Coefficients of the probability distribution.}
\textbf{a.} Coefficient $a_2$ in the case where $a_4$ is left free (blue disks, same as in \figref{fig:fig3}) and in the case where $a_4$ is set to zero for $u \geq 40$ (grey squares).
\textbf{b.} Fitted value of $a_4$, in log scale.
\textbf{c.} Ratio $a_2^2/a_4$.
Non-Gaussian signatures are expected when $a_2^2/a_4 \lesssim 1$.
In all panels, the dashed lines correspond to the asymptotic behaviours $u \ll \uc$ and $u \gg \uc$ (panel (\textbf{a}) only).
Note that in the disordered phase ($a_2 \geq 0$), the fitting of $a_4$ is unreliable because the measured distribution does not allow to easily distinguish between a quadratic and a quartic form.
}
\label{fig:figS8}
\end{figure}

In the main text, the measured probability distribution functions are fitted with the ansatz $\ee^{-(a_0 + a_2 |\psi|^2 + a_4 |\psi|^4)}$.
We show in \figref{fig:figS8}b the fitted values for the coefficient $a_4$. 
The form used to fit the experimental data is expected to be valid close to the transition point, and it is not granted that it remains valid away from the transition. 

In the specific case of the Mott insulator phase whose statistics in momentum space is well known \cite{carcy:2019a}, we expect a distribution whose quartic term is equal to zero ($a_4=0$).
As a matter of fact, when $u \geq 40$ the shape of the curves in the disordered phase does not allow to easily distinguish between a quadratic and a quartic form, leading to large correlations between the fitted $a_2$ and $a_4$ as well as large errorbars displayed in \figref{fig:figS8}a (blue disks, see also \figref{fig:fig3}c). 

Figure~\ref{fig:figS8}a (grey squares) shows the fit results obtained by enforcing $a_4=0$ for $u \geq 40$.
Under this assumption, we note that the fitted values for $a_2$ deep in the normal (Mott) phase ($u\gg \uc$) better match the expected value from the Mott statistics (dashed line).

In all cases, the coefficient $a_0$ is not fitted but fixed by normalisation of the probability by
\begin{equation}
    \sum_{n \in \mathbb{N}} \ee^{-(a_0 + a_2 |\psi|_n^2 + a_4 |\psi|_n^4)} = 1,
\end{equation}
where $|\psi|_n = \sqrt{n}$ are the values sampled by $\sqrt{N_0}$ experimentally.

\subsection{Critical probability distribution functions \label{SM:PDF}} 

In this section, we discuss the Probability Distribution Functions (PDFs) of critical systems and contrast them with those obtained from the Landau theory of critical phenomena.
Even though the system of interest is quantum, the transition we are studying here occurs at finite temperature. Therefore,  it can be described by an effective classical theory when focusing on large system sizes and large-scale properties \cite{sachdev:2011}.
We therefore focus on the PDF of classical critical systems.

In the case of a classical system with U(1) symmetry, the microscopic configurations are described in terms of a complex field $\phi(\bm{r})$, which can be defined on a lattice or in the continuum. Its thermodynamics is described by a Hamiltonian $\mathcal H [\phi(\bm{r})]$ and the Boltzmann weight
\begin{equation}
    P[\phi(\bm{r})] \propto \ee^{-\beta \mathcal H [\phi(\bm{r})]},
\end{equation}
which gives the probability of a configuration of the microscopic field. We assume that a control parameter $u$ drives the transition at some critical value $\uc$.

The physics of the system is usually more conveniently expressed in terms of a coarse-grained order-parameter field $\psi(\bm r) \in \mathbb{C}$ (which can, for instance, correspond to the spatial average of $\phi(\bm{r})$ over a mesoscopic region). The probability distribution of a configuration $\psi(\bm{r})$ is then given by a Ginzburg-Landau (GL) functional
\begin{equation}
    P_{\rm GL}[\psi(\bm{r})] \propto \ee^{-\mathcal F_{\rm GL} [\psi(\bm{r})]},
    \label{eq:probaGL}
\end{equation}
with 
\begin{equation}
\mathcal F_{\rm GL} [\psi(\bm{r})]=\int \dd\bm{r} \left[|\nabla \psi(\bm{r})|^2+\tilde a_2 |\psi(\bm{r})|^2+\tilde a_4|\psi(\bm{r})|^4+\ldots\right].
\label{eq:FGLSM}
\end{equation}
Working in the vicinity of the phase transition, we can assume that $\psi(\bm r)$ has a small amplitude and varies slowly in space, so that it is legitimate to expand the functional in powers of the field and its gradients.
Since the Ginzburg-Landau functional is an analytic function of $\psi(\bm{r})$ and $\psi^*(\bm{r})$ and is invariant under $U(1)$ transformations (i.e., phase rotations) of the field, only even powers of $|\psi(\bm{r})|$ and $|\nabla \psi(\bm{r})|$ are allowed in its expansion.

Focusing on the spatially-averaged/zero-momentum field $\psi=V^{-1/2}\int \dd\bm{r}\,\psi(\bm{r})$ (with $V$ the volume of the system), its statistical average $\braket{ \psi }$, taken using $P_{\rm GL}[\psi(\bm{r})]$,  corresponds to the order parameter that discriminates between different phases. 
The statistical properties of $\psi$ are given in terms of its PDF $P(\psi)$, which is the marginal distribution of $P_{\rm GL}$ after integrating out the modes with non-zero momenta.
We will use the common abuse of terminology and refer to $\psi$ as the order parameter.

Since all momentum modes interact with each other through the non-linear coupling (approximated by a quartic term in Eq.~\eqref{eq:FGLSM}), the computation of $P(\psi)$ from $P_{\rm GL}[\psi(\bm{r})]$ is a priori highly non-trivial.
Landau's theory of phase transitions amounts to performing a mean-field calculation and approximating $P(\psi)$ by
\begin{equation}
    P(\psi) \simeq P_{\rm L}(\psi) = e^{-V \mathcal{F}_{\rm L}(\psi)}.
    \label{eq:probaL}
\end{equation}
with $\mathcal{F}_{\rm L}(\psi)=\tilde  a_0+V\tilde a_2|\psi|^2+V^2\tilde a_4|\psi|^4$ the Landau free energy ($\tilde a_0$ is related to the normalization of $P_{\rm L}(\psi)$).
In the thermodynamic limit, $V\to \infty$, one finds that $|\braket{\psi}_{\rm L}|^2 \propto -\tilde a_2/\tilde a_4$ if $\tilde a_2<0$ and $|\braket{\psi}_{\rm L}|^2=0$ otherwise, recovering the standard Landau phenomenology.
Note that $\tilde a_2$ depends on $u$, but will generically vanish (linearly) at a value $u=u_{\rm L}\neq \uc$ where $\uc$ is the exact critical point. In this approximation and for a finite system, $\psi$ will have non-Gaussian statistics for a sufficiently small $\tilde a_2$ (such that $\tilde a^2_2 V / \tilde a_4 \lesssim 1$).

However, it is well known that in dimensions below four, the above mean-field picture breaks down. The coupling between small and large scales (or, alternatively, large- and small-momentum modes of $\psi(\bm{r})$) strongly renormalises the PDF, which, near the critical point, takes the scaling form
\begin{equation}
    P(\psi) = L^{2\beta/\nu-3} \tilde P (L^{\beta/\nu-3/2}\psi,L^{1/\nu}  \delta u).
    \label{eq:proba_scaling}
\end{equation}
Here, $L$ is the linear system size ($V=L^3$), $\beta$ and $\nu$ are the order-parameter and correlation-length critical exponents, respectively, and $ \delta u=(u-\uc)/u_c$ is the distance to the critical point.

The scaling function $\tilde P(x,y)$ is universal, in the sense that it does not depend on microscopic details of the system. These details enter only in so-called non-universal amplitudes, which amount to fixing the units of the variables $x$ and $y$ (for instance, the amplitude of the field $\psi$, and the link between $\delta u$ and the correlation length). The scaling function $\tilde P$ depends on the symmetry of the order parameter (here $U(1)$), the dimensionality of space (here $d=3$), and, importantly, the boundary conditions~\cite{binder:1981}. The critical exponents also depend on the former two, but not on the latter. 

Independently of the universality class, the shape of $\tilde P(x,y)$ is qualitatively of the following form \cite{balog:2022,rancon:2025}: i) for large $x$, it takes the form of a compressed exponential, $\ee^{-c(y) x^{\delta+1}}$, with $\delta$ the isothermal critical exponent (that depends on $\beta$ and $\nu$, with $\delta$ the critical exponent that relates the order parameter to the external field at criticality; $\delta\simeq 4.78 $ for the 3D XY universality class); ii) at small $x$, it can be approximated by $\ee^{-a_2 x^2-a_4 x^4-\ldots}$, where $a_2$ depends on $y$ and thus on $\delta u$. Importantly, $a_2$ changes sign at some $y_0\neq 0$ that depends on the boundary conditions. At fixed system size, this change of sign corresponds to $u=\uz\equiv\uc+y_0 L^{-1/\nu}$, implying that $\uz$ is generically different from $\uc$ (but $\uz\to\uc$ as $L\to\infty$). On the other hand, we expect $a_4$ to weakly depend on $\delta u$.
The region where $|\psi|$ assumes small values is therefore well captured by a naive Landau form Eq.~\eqref{eq:probaL} with $\tilde a_i\to a_i$ ($i=2,4$), even though the parameters and critical scaling are very different. Only the behaviour in the tail, through the exponent $\delta$, discriminates between the different universality classes.

It is important to stress that, as usual when discussing phase transitions, the limits $L\to\infty$, $\delta u\to0$, etc. do not commute. In particular, in the thermodynamic limit ($L\to\infty$ at fixed $\psi$ and $\delta u$), for all $\delta u>0$, $\xi\ll L$ and the central limit theorem holds: the PDF is thus Gaussian. To observe a universal non-Gaussian PDF, one needs to take the limit $L\to \infty$ (in practice, $L\gg \ell_G$) and $\delta u\to0$, such that $L/\xi$ is constant and of order one.

\subsection{Comparison of experimental distributions with 3D XY simulations and Landau predictions} 
\label{SM:PDF3DXY}

In this section, we compare the experimental PDF of the order parameter with the PDF predicted by a Landau-like approximation, and by Monte Carlo (MC) simulations of the homogeneous classical 3D XY model
\begin{equation}
H = -J \sum_{\langle \bm i, \bm j \rangle} \cos(\phi_{\bm i} - \phi_{\bm j}) 
\end{equation}
where $\phi_{\bm i}$ are phase variables, defined on a cubic lattice with either free or periodic boundary conditions. 
For this model the order parameter can be defined (in analogy with the bosonic system) as $\psi = L^{-3/2} \sum_{\bm i} \ee^{\ii \phi_{\bm i}}$; and the transition is controlled by the temperature.
The Landau-like approximation is $P(|\psi|=x) \propto e^{- (a_2 x^2 + a_4 x^4)}$. The MC simulations are performed using Wolff's algorithm, sampling more than a million order-parameter configurations. We have taken a sufficiently large system to ensure convergence of the PDFs with size. They obey the scaling form and behaviour discussed in Sec.~\ref{SM:PDF}. Note in particular that even though their shapes look qualitatively similar, they are in fact different, especially when compared at the same value of $L^{1/\nu}\delta u$ (with $\delta u$ being related to the inter-spin coupling in this context).

\begin{figure}[!t]
\centering
\includegraphics{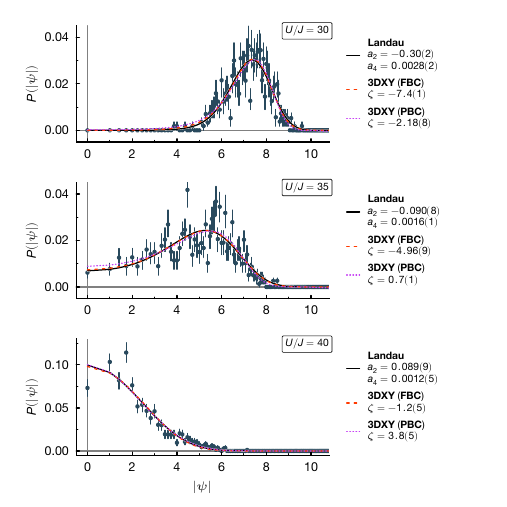}
\caption{
\textbf{Comparison of experimental FCS of the order parameter with various theories.}
Blue: experimental FCS, the errorbars represent the 68\%-confidence interval obtained from bootstrap method. Purple: fit by Landau-like expansion (see text). Red: fit from numerics of the 3D XY universality class with free boundary conditions (FBC). Orange: fit from numerics of the 3D XY universality class with periodic boundary conditions (PBC). For the latter two, we optimised over $L^{1/\nu}\delta u$ (here parametrised in terms of $\zeta\propto {\rm sign}(\delta u)L |\delta u|^{\nu}$) to obtain the best fit to the experimental data.
}
\label{fig:figS6}
\end{figure}

\figref{fig:figS6} shows the experimental data across the transition, compared to the best fit to a Landau-like distribution, and to the PDF of the 3D XY model at the distance to the critical point that gives the best overlap with the experimental PDF.
We observe that the three PDFs are quite similar, and that the differences between them cannot be distinguished experimentally with our signal-to-noise ratio (using $\sim 1000$ shots). Given the finite statistics of the experiments, in particular the lack of data for rare events, we find that the Landau-like approximation of the PDF works very well. The compressed exponential behaviour expected for rare events in the exact distribution only appears in the tail of the distribution itself, and its observation would require in principle a much larger number of experimental shots than that used in this work. The finite size of our sample introduces another fundamental limit to the reconstruction of the tail of the distribution.

The three classical theoretical predictions shown in \figref{fig:figS6} are continuous PDFs, while we measure discrete PDFs associated with the amplitude of a quantum field, whose square is the discrete atom number in the condensate mode. 
This leads to an issue regarding the comparison between the theoretical and the experimental PDFs. To resolve this issue, we normalise the continuous PDFs $f$ such that, $\sum_{n \in \mathbb{N}} f(|\psi|_n) = 1$, where $|\psi|_n = \sqrt{n}$ are the values sampled by $\sqrt{N_0}$ experimentally.

\subsection{Cumulants for a trapped classical field theory on a lattice}
\label{s.trap}

In the previous section, we have used the classical 3D XY model as a reference for the distribution of the order parameter. 
Yet this model lacks an important aspect of the experiment: the presence of a harmonic trap, shaping the density profile, and hence the strengths of the couplings between the phases of the bosonic fields in the system. 
In order to capture this effect, we studied a classical lattice field theory with a local complex field $\psi_{\bm i} = \sqrt{n_{\bm i}} \ee^{\ii \phi_{\bm i}}$ with a discrete amplitude ($n_{\bm i}$ being an integer), and subject to a parabolic trap.
Its Hamiltonian reads
\begin{eqnarray}
H &=& -J\sum_{\langle \bm i, \bm j \rangle} \left ( \psi_{\bm i}^* \psi_{\bm j} + {\rm c.c.} \right ) \nonumber \\
 &+& \sum_{\bm i} \left ( \frac{U}{2} |\psi_{\bm i}|^4 + V_{\bm i} |\psi_{\bm i}|^2 \right ), 
\end{eqnarray}%
where $U$ is the interaction parameter and $V_{\bm i} = V |{\bm i} - {\bm i}_0|^2$ the parabolic trapping potential. 
Defining the complex-valued order parameter as $\psi = L^{-3/2} \sum_{\bm i} \psi_{\bm i}$, we have sampled its distribution using Monte Carlo in the canonical ensemble --- namely for a fixed $N = \sum_{\bm i} n_{\bm i}$ --- across the thermal transition of the system. 
Different trapping potentials $V$ and particle numbers $N$ have been chosen so as to have an average density $\langle n_{\bm i} \rangle \approx 2$ at the centre, and the simulation boxes have been chosen in size such that the harmonically trapped gas is essentially insensitive to their boundaries.

\begin{figure*}[ht!]
\begin{center}
\includegraphics[scale=1]{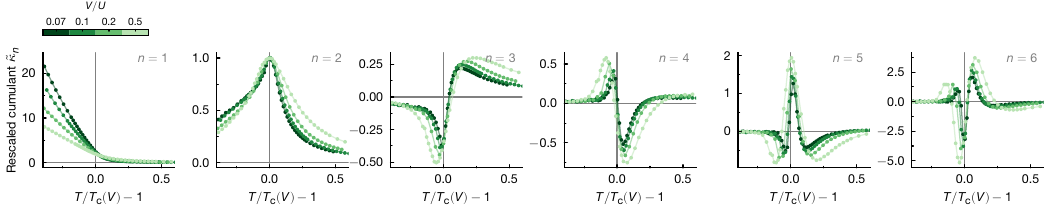}
\caption{
\textbf{Simulations with a trapped classical field theory on a lattice.} 
Cumulants up to order $n = 6$ computed using a classical field theory on a cubic lattice in the presence of harmonic trapping.
The datasets correspond to the following parameters $N=1000, V/U=0.5,  L=16$; $N=3000, V/U=0.2,  L=24$; $N=7000, V/U=0.1,  L=32$; $N=14000, V/U=0.07,  L=40$.}
\label{fig:figS11}
\end{center}
\end{figure*}

The order-parameter cumulants for the trapped field theory are shown in Fig.~\ref{fig:figS11}. 
We compare different trap and system sizes by defining a trap-dependent critical temperature $T_\mathrm{c}$ as the temperature at which the second-order cumulant $\kappa_2$ is maximum, and shifting the temperature axes in a trap-dependent fashion so as to align those peaks. Note that we do not attempt to collapse the curves as this would require a more advanced analysis \cite{campostrini:2009,campostrini:2010}, which goes beyond the scope of this work. 

Our main observation is that, regardless of the system size, the variation of the higher-order cumulants of the classical trapped system aligns with the universal ones of the 3D XY transition in homogeneous systems with periodic boundary conditions, shown in the insets of Fig.~\ref{fig:fig4} in the main text. In addition, it clearly differs from what is observed in the experiment. Hence we conclude that the harmonic trap cannot be held responsible for the discrepancies between the observed experimental cumulants and those reconstructed for classical systems. The origin of these discrepancies should rather be searched for in the quantum nature of the experimental system, as we shall discuss in the next section.

\subsection{Finite-size scaling of 3D quantum rotors.}

\subsubsection{Hardness of the numerical reconstruction of order-parameter statistics for the Bose-Hubbard model}

The results of the previous section have led us to the conclusion that classical models for the 3D XY transition cannot reproduce the behaviour of the higher-order cumulants observed in the experiment. This would incite us to reconstruct the order-parameter statistics directly for the Bose-Hubbard model, using quantum Monte Carlo (QMC), which allows for the unbiased calculation of thermal equilibrium properties of the system \cite{wessel:2004,carcy:2021}.
Using QMC formulated in the basis of the Fock states in position space, $|\{ n_{\bm i}\}\rangle$, one can directly calculate the first moment of the order parameter statistics $\langle N_0 \rangle$ using well established off-diagonal estimators based on the so-called worm (or directed-loop) algorithm. With some extra effort (\emph{i.e.}, using double worm algorithms) one can access the second moment as well \cite{roscilde:2008,fang:2016}. 
Yet reconstructing the $n$-th moment of the PDF $P(N_0)$ would require an $n$-worm algorithm, which is based on sampling increasingly rare events the higher the order $n$.

An apparent alternative to these difficulties is offered by switching to QMC formulated in the basis of  Fock states in quasi-momentum space, $|\{ n_{\bm q}\}\rangle$. Within this approach the populations would be simply sampled by the configuration updates, without the need for sophisticated estimators. The trouble is that, within the momentum basis, the (repulsive) interaction term takes the form $U/2 \sum_{\bm k, \bm k', \bm q} b^\dagger_{\bm k+\bm q} b^\dagger_{\bm k'-\bm q} b_{\bm k} b_{\bm k'}$, namely it is positive definite, largely off-diagonal, and it connects quartets of quasi-momenta without any sublattice structure. Under these circumstances, the interaction terms leads to negative weights for several path-integral configurations, or stochastic-series expansion terms ---  namely to a sign problem in the QMC approach.

Hence reconstructing high-order moments of the $P(N_0)$ distribution via QMC for the Bose-Hubbard model appears as a rather arduous enterprise. Nonetheless the full statistics of the order-parameter fluctuations can be efficiently sampled numerically in a related model, namely for quantum rotors, which offer a unique route to probing quantum effects on these fluctuations in an unbiased manner. 

\subsubsection{Quantum-rotor model} 

To reconstruct the non-Gaussian statistics of critical fluctuations in a quantum-mechanical model exhibiting the superfluid-to-normal-gas transition, we have performed simulations of the 3D quantum-rotor model.
The quantum-rotor model is the limit of the Bose-Hubbard model for large, integer filling $n_{\rm QR} \gg 1$ \cite{wallin:1994}. 
Decomposing the bosonic operator in terms of amplitude and phase, $\hat a_{\bm i} = \ee^{\ii\hat \phi_{\bm i}} \sqrt{\hat n_{\bm i}}$, with $[\hat n_{\bm i}, \hat \phi_{\bm i}] = \ii$, and rewriting the density as $\hat n_{\bm i} = n_{\rm QR} + \delta n_{\bm i}$, if  $\sqrt{\langle (\delta n_{\bm i})^2 \rangle} \ll n_{\rm QR} $  then one can write
\begin{equation}
\hat a_{\bm i}^\dagger \hat a_{\bm j} + {\rm h.c.} \approx 2 n_{\rm QR} \cos(\phi_{\bm i} - \phi_{\bm j})
\label{e.approx}
\end{equation}
upon neglecting density fluctuations. Density-density interactions introduce a quantum kinetic term in the Hamiltonian, since  $n_{\bm i}^2 = -\frac{\partial^2}{\partial \phi_{\bm i}^2}$.
At integer filling, all linear terms in the density drop from the Hamiltonian, leading to the quantum rotor Hamiltonian
\begin{equation}
H_{\rm QR} = -2Jn_{\rm QR} \sum_{\langle \bm{i},\bm{j} \rangle}  \cos(\phi_{\bm i} - \phi_{\bm j}) - \frac{U}{2} \sum_{\bm i} \frac{\partial^2}{\partial \phi_{\bm i}^2}~.
\end{equation}
As done in Ref.~\cite{bureik:2025}, the momentum distribution of the quantum rotor model can be sampled via a coherent-state path-integral approach \cite{wallin:1994, sondhi:1997,roscilde:2016}, leading to the possibility of reconstructing the FCS of the condensate population $N_0 = \hat a_0^\dagger \hat a_0$. 

\begin{figure*}[ht!]
\begin{center}
\includegraphics[scale=1]{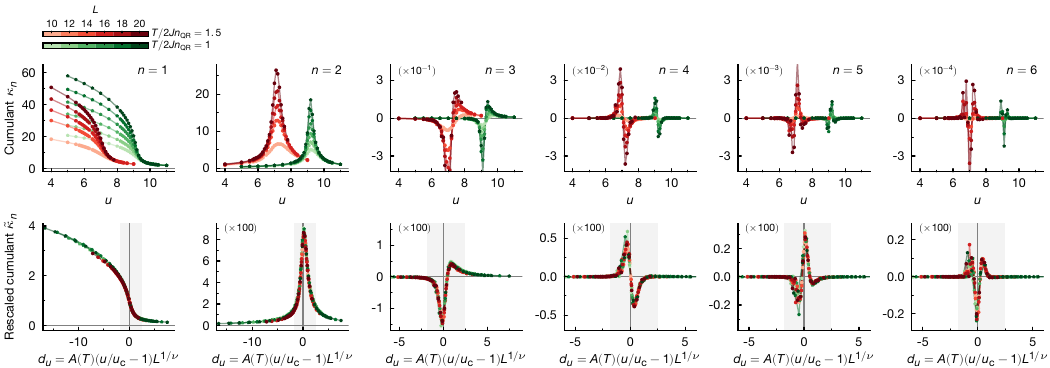}
\caption{
\textbf{Finite-size scaling analysis of 3D quantum rotors.}
Cumulants up to order $n = 6$ computed using the quantum rotor model.
The different colours represent different temperatures, $T/(2Jn_{\rm QR}) = 1.0$ (green) and $T/(2Jn_{\rm QR}) = 1.5$ (red).
The linear system size $L$ is encoded in the shading, ranging from $L = 10$ (lighter curves) to $L = 20$ (darker curves).
The top row shows the bare computation, for which the amplitude, position, and range of the critical regime depend both on temperature and system size.
For $n > 2$, the cumulants were rescaled as $\kappa_n \to  \kappa_n \times 10^{-n+2}$ to have convenient axis units.
The bottom row shows the result of the rescaling analysis, for which the cumulants collapse onto a universal curve (see text for details).
The shaded area represents roughly the region exhibiting non-Gaussian statistics, where $\kappa_n \neq 0$ at all orders, while the vertical line corresponds to $u=u_c$.
}
\label{fig:figS1}
\end{center}
\end{figure*}

\subsubsection{Universal scaling functions for cumulants at large temperatures} 

At finite temperature, the phase transition driven by increasing $U/J$ at fixed temperature belongs to the 3D XY universality class. From Eq.~\eqref{eq:proba_scaling}, we expect the cumulants of the order parameter to obey the finite-size scaling law
\begin{equation}
    \kappa_n(u,L,...) = L^{3n/2-n\beta/\nu} f_n(d_u,...),
\end{equation}
with $u = U/(2Jn_{\rm QR})$ the coupling constant, $d_u \propto (u/\uc - 1)\times L^{1/\nu}$, $\uc$ the (temperature-dependent) critical coupling, and the critical exponents $\beta$ and $\nu$  are taken as those of the 3D XY model, $\nu \simeq 0.6718$ and $\beta\simeq0.3485$. 
Here $f_n$ is a universal scaling function, which is uniquely characterized by the universality class and by global geometric properties (boundary conditions, aspect ratio of the dimensions, etc.). Its main argument is the control parameter of the transition, through the distance $d_u$ to the critical point. However, it possesses as well further arguments (indicated above as (...)), which can be neglected for sufficiently large sizes $L$ (see below for further discussion).

For a given temperature, the critical value $\uc$ is found by finding the point where all the different-$L$ curves $L^{\beta/\nu-3/2}\kappa_1(u,L)$  cross, since $L^{\beta/\nu-3/2}\kappa_1(\uc,L)=f_1(0)$ for all $L$. Then, for a given temperature and $n$, the curves $L^{n\beta/\nu-3n/2}\kappa_n(u,L)$ plotted as a function of $d_u$ are expected to collapse for all $L$ on a single, universal curve. 

However, to obtain the collapse of curves taken at different temperatures, one needs to fix two non-universal amplitudes, associated with the amplitude of the order parameter and to a length scale. 
Note that $\kappa_n$ has the dimension of the order parameter to the power $n$, which implies that $f_n$ has the dimension $\textrm{(order parameter)}^n \textrm{(length)}^{n\beta/\nu}$, and that $d_u$ has the dimension of a length to the power $1/\nu$. These two amplitudes could be obtained, for instance, from $f_1(0)$ and $f_1'(0)$, and using them to rescale all cumulants and the distance $u-u_c$ appropriately.
In practice, it is more convenient to define $\tilde\kappa_n = \kappa_n(u,L)/\kappa_1(\uc,L)^n$ and find $A(T)$ such that $\tilde\kappa_1$ plotted as a function of $d_u=A(T) \delta u L^{1/\nu}$ collapse on a single curve for all $T$. Then, all $\tilde\kappa_n$ plotted as a function of their corresponding $d_u$ also collapse on a single curve (which depends on $n$).

The curves shown in the inset of \figref{fig:fig4} correspond to the universal curves shown in \figref{fig:figS1}. The lines correspond to the average of the universal curves over the different sizes and temperatures. The shaded areas represent the dispersion of the different curves obtained with this method. The curves in the insets of \figref{fig:fig4} of the main text represent therefore the universal scaling functions (in the large $L$ limit) of the 3D XY transition with periodic boundary conditions and unit aspect ratios between different dimensions. 

Clearly, the experimental data also shown in \figref{fig:fig4} do not align with the predicted universal scaling functions for cumulants with $n \geq 4$. 
These deviations from the theoretical predictions also exhibit universal features, since they are found for two datasets with different geometries (transition occurring at the boundaries vs. at the centre of a trap)  --- see \figref{fig:fig5}. They cannot be attributed solely to the presence of a harmonic trap, since the data in Sec.~\ref{s.trap} show that cumulants of a classical trapped systems do not differ substantially from those of a homogeneous system (comparing \figref{fig:figS11} and \figref{fig:figS1}).

\subsubsection{Quantum corrections to scaling at low temperatures} 

We have shown above that we cannot reproduce the form of the cumulants measured in the experiment with \emph{any} classical model, nor with a quantum model in a regime of \emph{one-parameter} scaling (in the large $L$ limit). We are therefore led to suspect that our experimental observations have a quantum origin. More specifically, they could originate from quantum corrections to scaling that appear for moderate system sizes $L$ at finite temperature. 

In the thermodynamic limit of large system sizes $L$, quantum effects at finite temperature do not alter the nature of the phase transition, and therefore they do not alter the behaviour of the scaling functions. Nonetheless, they introduce a crossover length $\ell_T \sim (\hbar/k_B T)^{1/z}$ beyond which finite-temperature effects largely dominate over quantum contributions.  At any finite temperature, $\ell_T$ is finite and quantum contributions become irrelevant when $L \gg \ell_T$ as the large-scale behaviour of the transition extends beyond $\ell_T$: under this condition the classical scaling behaviour is recovered. On the other hand, with $L \lesssim \ell_T$, the contribution from quantum effects does not vanish and it is expected to affect the scaling behaviour \cite{sachdev:2011}. More specifically, the ratio $y=\ell_T/L$ should appear as an argument of the scaling functions, $f_n = f_n(x=d_u, y, ...)$.

\begin{figure*}[ht!]
\begin{center}
\includegraphics[scale=1]{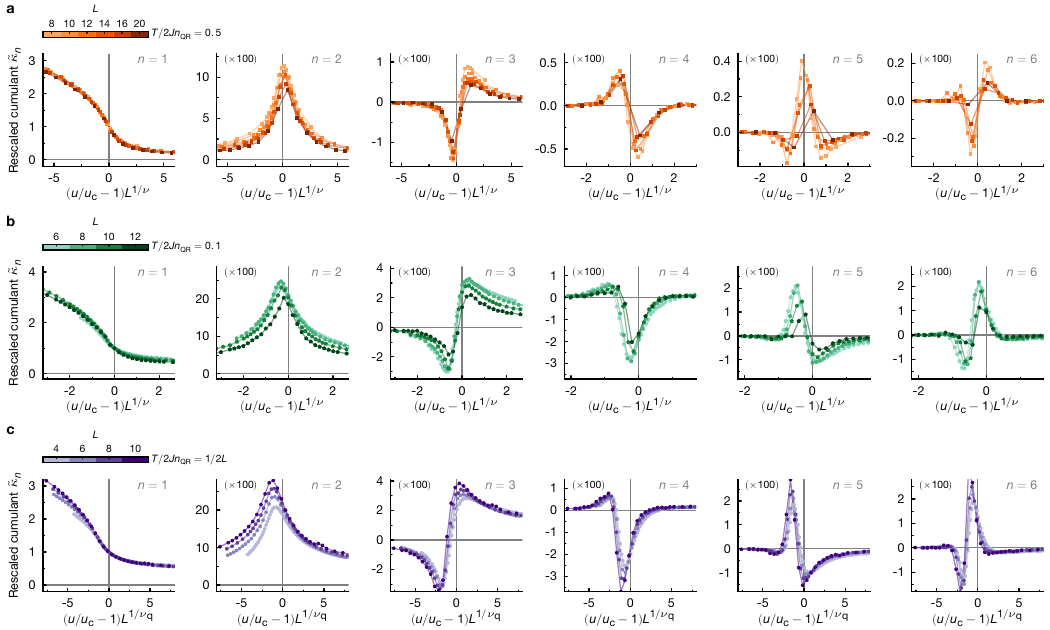}
\caption{
\textbf{Finite-size scaling analysis of 3D quantum rotors in the low-temperature regime.}
Cumulants up to order $n = 6$ computed using the quantum rotor model.
The different rows correspond to different temperatures, $T/(2Jn_{\rm QR}) = 0.5$ (top row, red), $T/(2Jn_{\rm QR}) = 0.1$ (middle row, green) and $T/(2Jn_{\rm QR}) = 1/2L$ (bottom row, violet).
The linear system size $L$ is encoded in the shading.
The numerical data on each row shows the result of the rescaling analysis, for which the cumulants collapse onto a universal curve (see text for details).
Note that the distance to the transition for the top and middle rows are rescaled using the critical exponent $\nu$ from the 3D XY universality class, while in the bottom row we use the exponent  $\nu_q=0.5$ associated to the 4D XY universality class of the quantum phase transition.
}
\label{fig:figS10}
\end{center}
\end{figure*}

A quantum correction to the scaling behaviour should therefore manifest itself as a difference between the $x$-dependence of the function $f_n(x,y,...)$ for $y \neq 0$ on the one hand; and that of $f_n(x,0,...)$ (the classical limit) on the other hand. To observe quantitatively these quantum corrections, we have considered the quantum-rotor model at increasingly large ratios $\ell_T/L$, by decreasing the (reduced) temperature $t=T/(2Jn_{\rm QR})$. When moving from $t=1$ and 1.5 studied in \figref{fig:figS1} to $t=0.5$ and 0.1, shown in \figref{fig:figS10}, we observe that, for the same system sizes, the emergence of a well-defined one-parameter scaling curve $f_n(x,0,...)$ deteriorates. This indicates corrections to scaling, namely the effect of a second, $y$ argument of the scaling function. 

Most remarkably, for the largest values of the $y$ argument (corresponding to the smaller system sizes and for the lowest temperatures) the universal scaling curves $f_n(x,y,...)$,  seen as functions of $x=d_u$ at fixed $y$,  exhibit shapes compatible with those measured in the experiment  ($n\geq 4$). 
This behaviour should be further enhanced when $L$ increases at a fixed ratio $\ell_T/L$, namely when the temperature is lowered upon increasing the system size.  Indeed, under these conditions, the transition under investigation is no longer the thermal superfluid-to-normal transition, but the quantum superfluid-to-Mott insulator transition. In the case of $T = 1/(2L)$, shown in \figref{fig:figS10}, a collapse of the cumulant data appears as the system size is increased upon using the critical exponents of the 4DXY Mott transition in 3D ($\beta = \nu = 1/2$). The corresponding scaling functions have similar shapes to those observed in the experiment.
  
\end{document}